\documentclass[runningheads]{llncs}

\usepackage{graphicx}
\usepackage{verbatim}

\usepackage{bussproofs}
\usepackage{amsfonts,amsmath,stmaryrd}
\usepackage{subfig}
\usepackage{wrapfig}
\usepackage{rotating}
\usepackage{adjustbox}
\usepackage{mdframed}
\usepackage{pdflscape}

\begin{document}

\title{Incorrectness Logic for Graph Programs}

\author{Christopher M. Poskitt}


\institute{Singapore Management University, Singapore\\
\email{cposkitt@smu.edu.sg}}

\maketitle 

\begin{abstract}
    Program logics typically reason about an over-approximation of program behaviour to prove the absence of bugs. Recently, program logics have been proposed that instead prove the \emph{presence} of bugs by means of \emph{under-approximate reasoning}, which has the promise of better scalability. In this paper, we present an under-approximate program logic for a nondeterministic graph programming language, and show how it can be used to reason deductively about program incorrectness, whether defined by the presence of forbidden graph structure or by finitely failing executions. We prove this `incorrectness logic' to be sound and complete, and speculate on some possible future applications of it.
    
\end{abstract}

\section{Introduction}

Many problems in computer science and software engineering can be modelled in terms of rule-based graph transformations~\cite{Heckel-Taentzer20a}, motivating research into verifying the correctness of grammars and programs based on this unit of computation. Various approaches towards this goal have been proposed, with techniques including model checking~\cite{Ghamarian-et_al12a}, unfoldings~\cite{Baldan-Corradini-Koenig08a,Koenig-Esparza10a}, $k$-induction~\cite{Schneider-Dyck-Giese20a}, weakest preconditions~\cite{Habel-Pennemann09a,Habel-Pennemann-Rensink06a}, abstract interpretation~\cite{Makhlouf-Percebois-Tran19a}, and program logics~\cite{Brenas-Echahed-Strecker18a,Poskitt13a,Poskitt-Plump12a}.


Verification approaches based on program logics and proofs typically reason about over-approximations of program behaviours to prove the absence of bugs. For instance, proving a partial correctness specification $\{pre\}P\{post\}$ guarantees that for states satisfying $pre$, every terminating execution of $P$ ends in a state satisfying $post$. Recently, authors have begun to investigate \emph{under-approximate} program logics that instead prove the \emph{presence} of bugs, motivated by the promise of better scalability that may result from reasoning only about the subset of paths that matter. De Vries and Koutavas~\cite{Vries-Koutavas11a} proposed the first program logic of this kind, using it to reason about state reachability for randomised nondeterministic algorithms. O'Hearn~\cite{OHearn20a} extended the idea to an \emph{incorrectness logic} that tracked both successful and erroneous executions. Under-approximate program logics have also been explored for local reasoning~\cite{Raad-et_al20a} and proving insecurity~\cite{Murray20a}.

An under-approximate specification $[pres] P [res]$ specifies a reachability property in the reverse direction: that every state satisfying $res$ (`result') is reachable by executing $P$ on \emph{some} state (not necessarily all) satisfying $pres$ (`presumption'). In other words, $res$ under-approximates the reachable states, allowing for sound reasoning about undesirable behaviours without any false positives, i.e.~a formal logical basis for bug catching. This is one of many dualities under-approximate program logics have with Hoare logics~\cite{Hoare69a}. Other important dualities include the inverted rule of consequence in which postconditions can be strenghtened (e.g.~by dropping disjuncts/paths), as well as the completeness proof which relies on \emph{weakest postconditions} rather than weakest preconditions.

In this paper, we present an under-approximate program logic for reasoning about the presence of bugs in nondeterministic attribute-manipulating graph programs. Following O'Hearn~\cite{OHearn20a}, we design it as an \emph{incorrectness logic}, and show how it can be used to reason deductively about the presence of forbidden graph structures or finitely failing executions (e.g.~due to the failure of finding a match for a rule). As our main technical result, we prove the soundness and relative completeness of our incorrectness logic with respect to a relational denotational semantics. The work in this paper is principally a theoretical exposition, but is motivated by some possible future applications, such as the use of incorrectness logic as a basis for sound reasoning in symbolic execution tools for graph and model transformations (e.g.~\cite{AlSibahi-Dimovski-Wasowski16a,Azizi-et_al20a,Oakes-et_al18a}).

The paper is organised as follows. In Section~\ref{sec:preliminaries} we provide preliminary definitions of graphs and graph morphisms. In Section~\ref{sec:programs_assertions} we define graph programs using a relational denotational semantics, as well as an assertion language (`E-conditions') for specifying properties of program states. In Section~\ref{sec:incorrectness_logic}, we present an incorrectness logic for graph programs and demonstrate it on some examples. In Section~\ref{sec:results}, we formally define the assertion transformations used in our incorrectness logic, and present our main soundness and completeness results. Finally, we review some related work in Section~\ref{sec:related_work} before concluding in Section~\ref{sec:conclusion}.

\section{Preliminaries}
\label{sec:preliminaries}

We use a definition of graphs in which edges are directed, nodes (resp.~edges) are partially (resp.~totally) labelled, and parallel edges are allowed to exist. All graphs in this paper will be totally labelled except for the interface graphs in rule applications (for technical reasons to support relabelling~\cite{Habel-Plump02a}).

A \emph{graph} over a label alphabet $\mathcal{C}$ is a system $G = \langle V_G, E_G, s_G, t_G, l_G, m_G \rangle$ comprising a finite set $V_G$ of \emph{nodes}, a finite set $E_G$ of \emph{edges}, \emph{source} and \emph{target functions} $s_G,t_G\!: E_G \rightarrow V_G$, a partial \emph{node labelling function} $l_G\!: V_G \rightarrow \mathcal{C}$, and a total \emph{edge labelling function} $m_G\!: E_G \rightarrow \{\square\}$. If $V_G = \emptyset$, then $G$ is the \emph{empty graph}, which we denote by $\emptyset$. Given a node $v\in V_G$, we write $l_G(v) = \bot$ to express that $l_G(v)$ is undefined. A graph $G$ is \emph{totally labelled} if $l_G$ is a total function. Note that for simplicity of presentation, in this paper, we label all edges with a `blank' label denoted by $\square$ and rendered as \adjincludegraphics[height=0.35\baselineskip,valign=c]{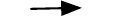} in diagrams. Note also that we use an undirected edge \adjincludegraphics[height=\baselineskip,valign=c]{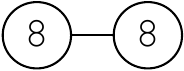} to represent a pair of edges \adjincludegraphics[height=\baselineskip,valign=c]{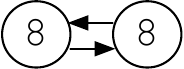}.

We write $\mathcal{G}(\mathcal{C}_\bot)$ (resp.~$\mathcal{G}(\mathcal{C})$) to denote the \emph{class} of all (resp.~all totally labelled) graphs over label alphabet $\mathcal{C}$. Let $\mathcal{L}$ denote the label alphabet $\mathbb{Z}^+$, i.e.~all non-empty sequences of integers. In diagrams we will delimit the integers of the sequence using colons, e.g.~5:6:7:8.

A \emph{graph morphism} $g\colon G\rightarrow H$ between graphs $G,H$ in $\mathcal{G}(\mathcal{C}_\bot)$ consists of two functions $g_V\colon V_G\rightarrow V_H$ and $g_E\colon E_G\rightarrow E_H$\/ that preserve sources, targets and labels; that is, $s_H\circ g_E=g_V\circ s_G$, $t_H\circ g_E=g_V\circ t_G$, $m_H \circ g_E = m_G$, and $l_H(g_V(v)) = l_G(v)$ for all nodes $v$ for which $l_G(v) \neq \bot$. We call $G,H$ respectively the \emph{domain} and \emph{codomain} of $g$.
	
A morphism $g$ is \emph{injective}\/ (\emph{surjective}) if $g_V$ and $g_E$ are injective (surjective). Injective morphisms are usually denoted by hooked arrows, $\hookrightarrow$. A morphism $g$ is an \emph{isomorphism} if it is injective, surjective, and satisfies $l_H(g_V(v)) = \bot$ for all nodes $v$ with $l_G(v) = \bot$. In this case $G$ and $H$\/ are \emph{isomorphic}, which is denoted by $G\cong H$. Finally, a morphism $g$ is an \emph{inclusion} if $g(x)=x$ for all nodes and edges $x$.

\section{Graph Programs and Assertions}
\label{sec:programs_assertions}

We begin by introducing the graph programs that will be the target of our incorrectness logic, as well as an assertion language (`E-conditions') that will be used for specifying properties of the program states (which consist of graphs). To allow for a self-contained presentation, our programs are a simplified `core' of full-fledged graph programming languages (e.g.~\textsc{GP 2}~\cite{Plump11a}) which have several more features for practicality (e.g.~additional types, negative application conditions).

First, we define the underlying unit of computation in graph programs: the application of a graph transformation rule with relabelling.

\begin{definition}[Rule]\rm
    A \emph{(concrete) rule} $r\!:\langle L\hookleftarrow K \hookrightarrow R \rangle$ comprises totally labelled graphs $L,R\in\mathcal{G}(\mathcal{L})$, a partially labelled graph $K\in\mathcal{G}(\mathcal{L}_\bot)$, and inclusions $K\hookrightarrow L$, $K\hookrightarrow R$. We call $L,R$ the \emph{left-} and \emph{right-hand graphs of $r$}, and $K$ its \emph{interface}.
    \qed
\end{definition}

Intuitively, an application of a rule $r$ to a graph $G\in\mathcal{G}(\mathcal{L})$ removes items in $L-K$, preserves those in $K$, adds the items in $R-K$, and relabels the unlabelled nodes in $K$. An injective morphism $g\!:L \hookrightarrow G$ is a \emph{match} for $r$ if it satisfies the dangling condition, i.e.~no node in $g(L)-g(K)$ is incident to an edge in $G-g(L)$. In this case, $G$ directly derives $H\in\mathcal{G}(\mathcal{L})$ with \emph{comatch} $h\!:R\hookrightarrow H$, denoted $G\Rightarrow_{r,g,h} H$ (or just $G\Rightarrow_{r} H$), by: (1):~removing all nodes and edges in $g(L)-g(K)$; (2)~disjointly adding all nodes and edges from $R-K$, keeping their labels (for $e \in E_R - E_K$, $s_H(e)$ is $s_R(e)$ if $s_R(e) \in V_R-V_K$, otherwise $g_V(s_R(e))$; targets analogous); (3)~for every node in $K$, $l_H(g_V(v))$ becomes $l_R(v)$. Semantically, direct derivations are constructed as two `natural pushouts' (see~\cite{Habel-Plump02a} for the technical details).

In practical graph programming languages, we need a more powerful unit of computation---the rule schema---which describes (potentially) infinitely many concrete rules by labelling the graphs over expressions. We define a simple abstract syntax `Exp' (Figure~\ref{fig:abstract_label_syntax}) which derives a label alphabet of (lists of) integer expressions, including variables (`Var') of type integer.

\begin{figure}[t]
\begin{center}
\renewcommand{\arraystretch}{1.2}
\begin{tabular}{lcl}
Exp & ::= & Integer $\mid$ Integer ':' Exp \\
Integer & ::= & Digit \{Digit\} $\mid$ Var $\mid$ '$\mathtt{-}$' Integer $\mid$ Integer ArithOp Integer \\
ArithOp & ::= & '$\mathtt{+}$' $\mid$ '$\mathtt{-}$' $\mid$ '$\mathtt{*}$' $\mid$ '$\mathtt{/}$'
\end{tabular}
\end{center}
	\caption{Abstract syntax of rule schema labels}
	\label{fig:abstract_label_syntax}
\end{figure}

A graph in $\mathcal{G}(\mathcal{L})$ can be obtained from a graph in $\mathcal{G}(\mathrm{Exp})$ by means of an \emph{interpretation}, which is a partial function $I\!:\mathrm{Var} \rightarrow \mathbb{Z}$. We denote the domain of $I$ by $\mathrm{dom}(I)$, and the set of variables used in a graph $G\in\mathcal{G}(\mathrm{Exp})$ by $\mathrm{vars}(G)$. If $\mathrm{vars}(G) \subseteq \mathrm{dom}(I)$, then $G^I\in\mathcal{G}(\mathcal{L})$ is the graph obtained by evaluating the expressions in the standard way, with variables $\mathtt{x}$ substituted for $I(\mathtt{x})$. Interpretations may also be applied to morphisms, e.g.~$p\!:P\hookrightarrow C$ becomes $p^I\!:P^I\hookrightarrow C^I$.

\begin{definition}[Rule schema]\rm
    A \emph{rule schema} $r\!: \langle L \Rightarrow R \rangle$ with $L,R\in \mathcal{G}(\mathrm{Exp})$ represents concrete rules $r^I\!: \langle L^I \hookleftarrow K \hookrightarrow R^I \rangle$ where $\mathrm{dom}(I) = \mathrm{vars}(L)$ and $K$ consists of the preserved nodes only (with all nodes unlabelled). Note that we assume for any rule schema, $\mathrm{vars}(R) \subseteq \mathrm{vars}(L)$.
    \qed
\end{definition}

The application of a rule schema $r = \langle L \Rightarrow R \rangle$ to a graph $G\in\mathcal{G}(\mathcal{L})$ consists of the following steps: (1)~choose an interpretation $I$ with $\mathrm{dom}(I) = \mathrm{vars}(L)$; (2)~choose a \emph{match}, i.e.~a morphism $g\!:L^I\hookrightarrow G$ that satisfies the dangling condition with respect to $r^I\!: \langle L^I \hookleftarrow K \hookrightarrow R^I\rangle$; (3)~apply $r^I$ with match $g$. If a graph $H$ with comatch $h\!:R^I\hookrightarrow H$ is derived from $G$ via these steps, we write $G\Rightarrow_{r,g,h}$ (or just $G\Rightarrow_r H$). Moreover, if a graph $H$ can be derived from a graph $G$ via some $r$ in a set of rule schemata $\mathcal{R}$, we write $G\Rightarrow_\mathcal{R} H$ (i.e.~nondeterministic choice of rule schema). If no rule schema in the set has a match for $G$, we write $G\not\Rightarrow_\mathcal{R}$ (i.e.~finite failure).

\begin{figure}[t]%
    \centering
    \subfloat{{\includegraphics[width=0.75\textwidth]{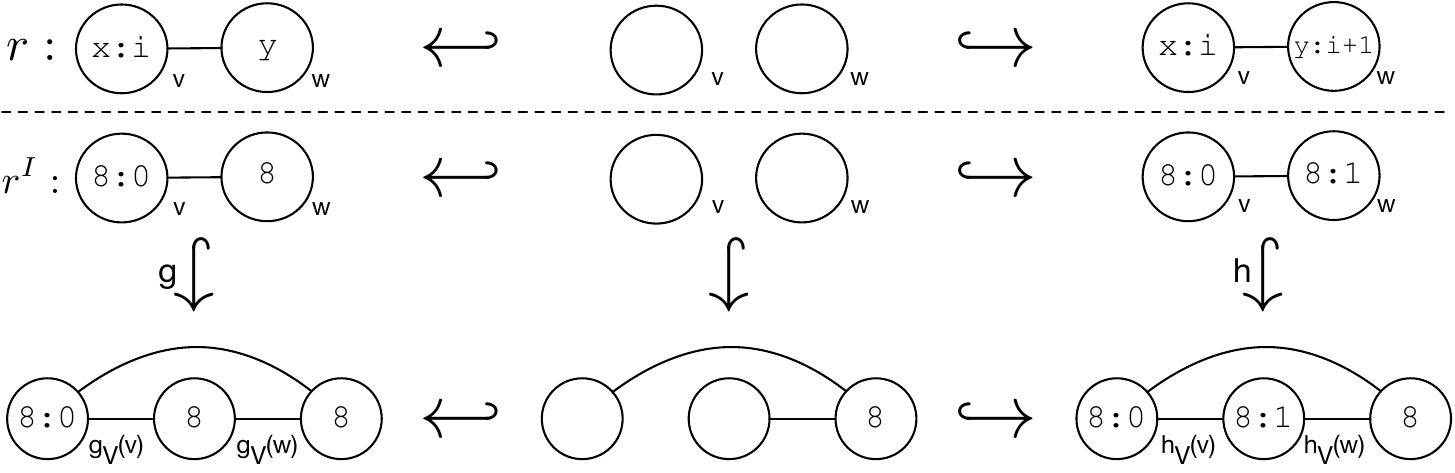} }}%
    \caption{Example rule schema application}
    \label{fig:rule_schema_application}
\end{figure}

\begin{example}[Rule schema application]
    Figure~\ref{fig:rule_schema_application} displays a rule schema $r\!: \langle L \hookleftarrow K \hookrightarrow R \rangle$ with its interface (top row), a possible instantiation $r^I$ where $I(\mathtt{x}) = I(\mathtt{y}) = 8$ and $I(\mathtt{i}) = 0$ (middle row). Finally, the bottom row depicts a direct derivation from $G$ (bottom left) to $H$ (bottom right) via $r^I$.
    \qed
\end{example}

\begin{definition}[Graph programs]\rm
    \emph{(Graph) programs} are defined inductively. Given a set of rule schemata $\mathcal{R}$, $\mathcal{R}$ and $\mathcal{R!}$ are programs. If $P,Q$ are programs and $\mathcal{R}$ a set of rule schemata, then $P;Q$ and $\mathtt{if}\ \mathcal{R}\ \mathtt{then}\ P\ \mathtt{else}\ Q$ are programs.
    \qed
\end{definition}

Intuitively, $\mathcal{R}$ denotes a single nondeterministic application of a rule schemata set. This results in failure if none of the rules are applicable to the current graph. The program $\mathcal{R}!$ denotes as-long-as-possible iteration of $\mathcal{R}$, in which the iteration terminates the moment that $\mathcal{R}$ is no longer applicable to the current graph (the program never fails). Finally, the program $P;Q$ denotes sequential composition, and $\mathtt{if}\ \mathcal{R}\ \mathtt{then}\ P\ \mathtt{else}\ Q$ denotes conditional branching, determined by testing the applicability of $\mathcal{R}$ (note that $\mathcal{R}$ will not transform the current graph).

Each graph program is given a simple relational denotational semantics (in the style of~\cite{OHearn20a}). We associate each program $P$ with two semantic functions, $\llbracket P \rrbracket ok$ and $\llbracket P \rrbracket er$, which respectively describe state (i.e.~graph) transitions for successful and finitely failing computations. Unlike operational semantics for graph programs (e.g.~\cite{Plump11a}), we do not explicitly track a `fail' state, but rather return pairs $(G,H)$ where $H$ is the last graph derived from $G$ before the failure.

\begin{definition}[Semantics]\rm
    The \emph{semantics} of a graph program $P$ is given by a binary relation $\llbracket P \rrbracket \epsilon \subseteq \mathcal{G}(\mathcal{L}) \times \mathcal{G}(\mathcal{L})$, defined according to Figure~\ref{fig:semantics}.
    \qed
\end{definition}

\begin{figure}[!t]
	{\footnotesize\begin{center}
	\begin{mdframed}
	\begin{tabular}{ r c l }

    $\llbracket \mathcal{R} \rrbracket ok$ &\ $=$ &\ $\{ (G, H) \mid G \Rightarrow_\mathcal{R} H \}$ \\
    
    $\llbracket \mathcal{R} \rrbracket er$ &\ $=$ &\ $\{ (G, G) \mid G \not\Rightarrow_\mathcal{R}  \}$ \\
    
    $\llbracket P;Q \rrbracket \epsilon$ &\ $=$ &\ $\{ (G, H) \mid \exists G'. (G, G') \in \llbracket P \rrbracket ok\ \mathrm{and}\ (G', H) \in \llbracket Q \rrbracket \epsilon \}  $ \\
    
    & &\  $\cup\ \left(\mathrm{if}\ \epsilon = er\ \mathrm{then}\ \{ (G,H) \mid (G, H) \in \llbracket P \rrbracket er \}\right)$ \\
    
    $\llbracket \mathcal{R}! \rrbracket ok$ &\ $=$ &\ $\llbracket \mathcal{R} \rrbracket er \cup \llbracket \mathcal{R};\mathcal{R}! \rrbracket ok$ \\
    
    
    $\llbracket \mathcal{R}! \rrbracket er$ &\ $=$ &\ $\emptyset$ \\
    
    $\llbracket \mathtt{if}\ \mathcal{R}\ \mathtt{then}\ P\ \mathtt{else}\ Q \rrbracket \epsilon$ &\ $=$ &\ $\{ (G, H) \mid \exists G'. (G,G') \in \llbracket \mathcal{R} \rrbracket ok\ \mathrm{and}\ (G,H) \in \llbracket P \rrbracket \epsilon \}$ \\
    
    & &\ $\cup\ \{ (G,H) \mid (G,G) \in \llbracket \mathcal{R} \rrbracket er\ \mathrm{and}\ (G,H) \in \llbracket Q \rrbracket \epsilon \}$

	\end{tabular}
	\end{mdframed}
	\end{center}
	}
	\caption{A relational denotational semantics for graph programs}\label{fig:semantics}
	\end{figure}

Note that divergence is treated in an implicit way: a program that always diverges is associated with empty relations. For example, $\llbracket \langle \emptyset \Rightarrow \emptyset \rangle ! \rrbracket ok = \emptyset$.

\begin{figure}[t]%
    \centering
    \subfloat{{\includegraphics[width=0.5\textwidth]{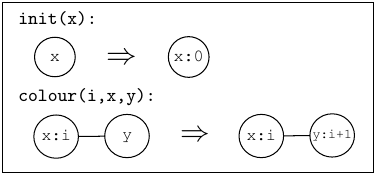} }}%
    \qquad
    \subfloat{{\includegraphics[width=0.38\textwidth]{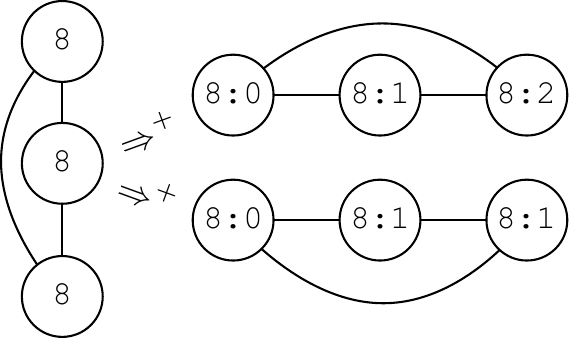} }}%
    \caption{Rules for the program \texttt{init; colour!} and two possible executions}
    \label{fig:buggy_program}
\end{figure}

\begin{example}[Buggy colouring]\rm
    Figure~\ref{fig:buggy_program} contains an example graph program $P = \mathtt{init;colour!}$ that purportedly computes a graph colouring, i.e.~an association of integers (`colours') with nodes such that no two adjacent nodes are associated with the same colour. The program nondeterministically assigns a colour of `0' to a node, encoding it as the second element of the label's sequence, before iteratively matching adjacent pairs of coloured/uncoloured nodes and assigning a colour to the latter obtained by incrementing the colour of the former. Note that the edges are undirected for simplicity.
    
    Two possible executions are shown in Figure~\ref{fig:buggy_program}, the first of which leads to a correct colouring, and the second of which leads to an illegal one. Moreover, the program can finitely fail on input graphs for which $\mathtt{init}$ has no match. (We shall use incorrectness logic to logically prove the presence of such outcomes.)
\end{example}

Before we can define an incorrectness logic for graph programs, we require an assertion language for expressing properties of the states, i.e.~graphs in $\mathcal{G}(\mathcal{L})$. For this purpose we shall use nested conditions with expressions (`E-conditions'), which allow for the specification of properties at the same level of abstraction, i.e.~by graph morphisms annotated with expressions. The concept of E-conditions was introduced in prior work~\cite{Poskitt13a,Poskitt-Plump12a}, but we shall present an alternative definition that more cleanly separates the quantification of graph structure and integer variables (the latter was handled implicitly in previous work, which led to more complicated assertion transformations).

\begin{definition}[E-condition]\rm
    Let $P$ denote a graph in $\mathcal{G}(\mathrm{Exp})$. A \emph{nested condition with expressions (\emph{short.}~E-condition) over $P$} is of the form $\mathtt{true}$, $\gamma$, $\exists\mathtt{x}. c$, or $\exists a. c'$, where $\gamma$ is an interpretation constraint (i.e.~a Boolean expression over `Exp'), $\mathtt{x}$ is a variable in Var, $c$ is an E-condition over $P$, $a\!:P\hookrightarrow C$ is an injective graph morphism over $\mathcal{G}(\mathrm{Exp})$, and $c'$ is an E-condition over $C$. Moreover, $\neg c_1$, $c_1 \wedge c_2$, and $c_1 \vee c_2$ are E-conditions over $P$ if $c_1,c_2$ are E-conditions over $P$.
    \qed
\end{definition}

The \emph{free variables} of an E-condition $c$, denoted $\mathrm{FV}(c)$, are those variables present in node labels and interpretation constraints that are not bound by any variable quantifier (defined in the standard way). If $c$ is defined over the empty graph $\emptyset$ and $\mathrm{FV}(c) = \emptyset$, we call $c$ an \emph{E-constraint}. Furthermore, a mapping of free variables to expressions $\sigma = \left(\mathtt{x}_1\mapsto e_1, \cdots\right)$ is called a \emph{substitution}, and $c^\sigma$ denotes the E-condition $c$ but with all free variables $\mathtt{x}$ substituted for $\sigma(\mathtt{x})$.

\begin{definition}[Satisfaction of E-conditions]\rm
    Let $c$ denote an E-condition over $P$, $I$ an interpretation with $\mathrm{dom}(I) = \mathrm{FV}(c)$, and $p\!:P^I\hookrightarrow G$ an injective morphism over $\mathcal{G}(\mathcal{L})$. The \emph{satisfaction} relation $p \models^I c$ is defined inductively.
    
    If $c$ has the form $\mathtt{true}$, then $p \models^I c$ always. If $c$ is an interpretation constraint $\gamma$, then $p \models^I c$ if $\gamma^I = \mathrm{true}$ (defined in the standard way). If $c$ has the form $\exists \mathtt{x}. c'$ where $c'$ is an E-condition over $P$, then $p \models^I c$ if $p\models^{I[\mathtt{x}\mapsto v]} c'$ for some $v\in\mathbb{Z}$. If $c$ has the form $\exists a\!:P\hookrightarrow C. c'$ where $c'$ is an E-condition over $C$, then $p \models^I c$ if there exists an injective morphism $q\!:C^I \hookrightarrow G$ such that $q \circ a^I = p$ and $q \models^I c'$.
    
    \begin{center}
        \includegraphics[width=0.2\textwidth]{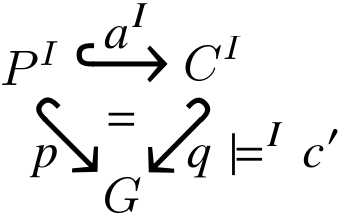}
    \end{center}
    
    Finally, the satisfaction of Boolean formulae over E-conditions is defined in the standard way.
    \qed
\end{definition}

The satisfaction of E-constraints by graphs is defined as a special case of the general definition. That is, a graph $G\in\mathcal{G}(\mathcal{L})$ \emph{satisfies} an E-constraint $c$, denoted $G \models c$, if $i_G\!:\emptyset\hookrightarrow G \models^{I_\emptyset} c$, where $I_\emptyset$ is the empty interpretation, i.e.~with $\mathrm{dom}(I_\emptyset) = \emptyset$.

For brevity, we write $\mathtt{false}$ for $\neg\mathtt{true}$, $c\Longrightarrow d$ for $\neg c \vee d$, $\forall\mathtt{x}.c$ for $\neg\exists\mathtt{x}.\neg c$, $\forall a.c$ for $\neg\exists a.\neg c$, and $\exists\mathtt{x}_1, \cdots \mathtt{x}_n.c$ for $\exists\mathtt{x}_1.\cdots \exists\mathtt{x}_n.c$ (analogous for $\forall$). Furthermore, if the domain of a morphism can unambiguously be inferred from the context, we write only the codomain. For example, the E-constraint $\exists \emptyset\hookrightarrow C.\ \exists C\hookrightarrow C'.\ \mathtt{true}$ can be written as $\exists C.\ \exists C'$.

\begin{example}[E-constraint]
    The following E-constraint expresses that for every pair of integer-labelled nodes, if the labels differ, then the nodes are adjacent:
    \[ \forall\mathtt{x},\mathtt{y}.\ \forall \adjincludegraphics[height=\baselineskip,valign=c]{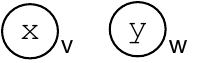}.\ \mathtt{x}\neq\mathtt{y} \Longrightarrow \exists \adjincludegraphics[height=\baselineskip,valign=c]{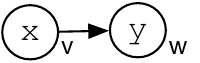} \vee \exists \adjincludegraphics[height=\baselineskip,valign=c]{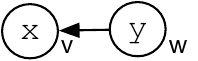} \]
    Note that $v,w$ are node identifiers to indicate which nodes are the same along the chain of nested morphisms, as can be seen when denoting them in full:
    \[ \forall\mathtt{x},\mathtt{y}.\ \forall \emptyset\hookrightarrow\adjincludegraphics[height=\baselineskip,valign=c]{fig_ex_ec_1.pdf}.\ \mathtt{x}\neq\mathtt{y} \Longrightarrow \exists \adjincludegraphics[height=\baselineskip,valign=c]{fig_ex_ec_1.pdf}\hookrightarrow\adjincludegraphics[height=\baselineskip,valign=c]{fig_ex_ec_2.pdf} \vee \exists \adjincludegraphics[height=\baselineskip,valign=c]{fig_ex_ec_1.pdf}\hookrightarrow\adjincludegraphics[height=\baselineskip,valign=c]{fig_ex_ec_3.pdf} \]
    \noindent These node identifiers may be omitted when the mappings are unambiguous.
\end{example}

\section{Proving the Presence of Bugs}
\label{sec:incorrectness_logic}

Before we define the proof rules of our incorrectness logic, it is important to define what an \emph{incorrectness specification} is and what it means for it to be \emph{valid}. In over-approximate program logics (e.g.~\cite{Poskitt13a,Poskitt-Plump12a}) a specification is given in the form of a triple, $\{c\}P\{d\}$, which under partial correctness expresses that if a graph satisfies precondition $c$, and program $P$ successfully terminates on it, then the resulting graph will always satisfy $d$. The postcondition $d$ over-approximates the graphs reachable upon termination of $P$ from graphs satisfying $c$.

Incorrectness logic~\cite{OHearn20a}, however, is based on under-approximate reasoning, for which a specification $[c]P[d]$ has a rather different meaning (and thus a different notation). Here, we call the pre-assertion $c$ a \emph{presumption} and the post-assertion $d$ a \emph{result}. The triple specifies that if a graph satisfies $d$, then it can be derived from \emph{some} graph satisfying $c$ by executing $P$ on it. In other words, $d$ under-approximates the states reached as a result of executing $P$ on graphs satisfying $c$. It does not specify that every graph satisfying $c$ derives a graph satisfying $d$, and it does not preclude graphs satisfying $\neg c$ from deriving such graphs either.

The principal benefit of proving such triples is then proving the \emph{presence of bugs}, and can be thought of as providing a possible formal foundation for static bug catchers, e.g.~symbolic execution tools. In graph programs, this amounts to formal proofs of the presence of \emph{illegal graph structure}, but it can also facilitate proofs of the presence of \emph{finite failure}. To accommodate this, we adopt O'Hearn's approach~\cite{OHearn20a} of tracking exit conditions $\epsilon$ in the result, $[c]P[\epsilon: d]$, using $ok$ to represent normal executions and $er$ to track finite failures.

\begin{definition}[Under-approximate validity]\label{defn:validity}\rm
    Let $c,d$ denote E-constraints, $P$ a graph program, and $\epsilon$ an exit condition. A specification $[c]\ P\ [\epsilon: d]$ is \emph{valid}, denoted $\models [c]\ P\ [\epsilon: d]$, if for every graph $H\in\mathcal{G}(\mathcal{L})$ such that $H \models d$, there exists a graph $G\in\mathcal{G}(\mathcal{L})$ such that $G\models c$ and $(G,H) \in \llbracket P \rrbracket \epsilon$.
    \qed
\end{definition}

Figure~\ref{fig:incorrectness_logic} presents the axioms and proof rules of our incorrectness logic for graph programs, which are adapted from O'Hearn's incorrectness logic for imperative programs~\cite{OHearn20a}. We say that a triple is \emph{provable}, denoted $\vdash [c]P[\epsilon: d]$, if it can be instantiated from any axiom, or deduced as the consequent of any proof rule with provable antecedents. We use the notation $\vdash [c]P[ok: d_1][er: d_2]$ as shorthand for two separate triples, $\vdash[c]P[ok:d_1]$ and $\vdash[c]P[er: d_2]$.

	\begin{figure}[!t]
	{\footnotesize\begin{center}
	\begin{mdframed}
	\begin{tabular}{ p{0.5\textwidth} p{0.5\textwidth} }
	
	\multicolumn{2}{p{\textwidth}}{
    \vspace{-10pt}
    \begin{prooftree}
	\AxiomC{\textsc{RuleSetSucc}\ \ $\vdash [ c \wedge \mathrm{App}(\mathcal{R}) ]~ \mathcal{R} ~ [ ok: \mathrm{WPost}(\mathcal{R}, c) ][er: \mathtt{false}]$}
	\end{prooftree}
	}\\[-15pt]
	
	\multicolumn{2}{p{\textwidth}}{
	
	\begin{prooftree}
	\AxiomC{\textsc{RuleSetFail}\ \ $\vdash [ c \wedge \neg\mathrm{App}(\mathcal{R}) ]~ \mathcal{R} ~ [ok: \mathtt{false}][ er: c \wedge \neg\mathrm{App}(\mathcal{R}) ]$}
	\end{prooftree}
	}\\[-15pt]

	\begin{prooftree}
	\AxiomC{$\vdash [ c ]~ P~ [ok: e ]$}
	\AxiomC{$\vdash [ e ]~ Q~ [ \epsilon: d ]$}
	\LeftLabel{\textsc{SeqSucc}}
	\BinaryInfC{$\vdash [ c ]~ P\mathtt{;}~ Q ~ [ \epsilon: d ]$}
	\end{prooftree}

	&
	\begin{prooftree}
	\AxiomC{$\vdash [ c ]~ P~ [er: d ]$}
	\LeftLabel{\textsc{SeqFail}}
	\UnaryInfC{$\vdash [ c ]~ P\mathtt{;}~ Q ~ [ er: d ]$}
	\end{prooftree}
	\\[-15pt]

	\multicolumn{2}{p{\textwidth}}{

	\begin{prooftree}
	\AxiomC{$\vdash [ c \wedge \mathrm{App}(\mathcal{R}) ]~ P~ [\epsilon: d ]$ or $\vdash [ c \wedge \neg\mathrm{App}(\mathcal{R}) ]~ Q~ [ \epsilon: d ]$}
	\LeftLabel{\textsc{IfElse}} 
	\UnaryInfC{$\vdash [ c ]~ \mathtt{if}\ \mathcal{R}\ \mathtt{then}\ P\ \mathtt{else}\ Q~ [ \epsilon: d ]$}
	\end{prooftree}
	
	}\\[-15pt]
	
	\multicolumn{2}{p{\textwidth}}{

	\begin{prooftree}
	\AxiomC{$c \Longleftarrow c'\ \ \vdash [ c' ]~ P~ [ \epsilon: d' ]\ \ d' \Longleftarrow d$}
	\LeftLabel{\textsc{Cons}} 
	\UnaryInfC{$\vdash [ c ]~ P ~ [ \epsilon: d ]$}
	\end{prooftree}
	
	}\\[-15pt]
	
	\multicolumn{2}{p{\textwidth}}{

    \begin{prooftree}
	\AxiomC{\textsc{IterZero}\ \ $\vdash [ c \wedge \neg\mathrm{App}(\mathcal{R}) ]~ \mathcal{R}! ~ [ ok: c \wedge \neg\mathrm{App}(\mathcal{R}) ][er: \mathtt{false}]$}
	\end{prooftree}
	}\\[-15pt]
	
	\multicolumn{2}{p{\textwidth}}{

	\begin{prooftree}
	\AxiomC{$\vdash [ c \wedge \mathrm{App}(\mathcal{R}) ]~ \mathcal{R}; \mathcal{R}!~ [ok: d \wedge \neg\mathrm{App}(\mathcal{R}) ]$}
	\LeftLabel{\textsc{Iter}}
	\UnaryInfC{$\vdash [ c \wedge \mathrm{App}(\mathcal{R}) ]~ \mathcal{R}! ~ [ ok : d \wedge \neg\mathrm{App}(\mathcal{R}) ]$}
	\end{prooftree}
	
	}\\[-15pt]
	
	\multicolumn{2}{p{\textwidth}}{

	\begin{prooftree}
	\AxiomC{$\vdash [ c_{i-1} ]~ \mathcal{R}~ [ ok: c_i  ]$ for all $0 < i \leq n$, and $c_n \Longrightarrow \neg\mathrm{App}(\mathcal{R})$}
	\LeftLabel{\textsc{IterVar}}
	\UnaryInfC{$\vdash [ c_0 ]~ \mathcal{R}!~ [ ok: c_n ]$}
	\end{prooftree}

	}

	\end{tabular}
	\end{mdframed}
	\end{center}
	}
	\caption{Incorrectness axioms and proof rules for graph programs}\label{fig:incorrectness_logic}
	\end{figure}

Note that a number of axioms and proof rules rely on some transformations that we have not yet defined: $\mathrm{App}(\mathcal{R})$, which expresses the existence of a match for $\mathcal{R}$, and $\mathrm{WPost}(\mathcal{R}, c)$, which expresses the \emph{weakest postcondition} that must be satisfied to guarantee the existence of a pre-state satisfying $c$. These transformations will be formally defined in Section~\ref{sec:results}.

The axioms \textsc{RuleSetSucc} and \textsc{RuleSetFail} allow for reasoning about the most fundamental unit of graph programs: rule schema application. The former covers the successful case: if a graph satisfies the weakest postcondition for rule schemata set $\mathcal{R}$ and E-constraint $c$, then it can be derived from some graph satisfying the presumption $c \wedge \mathrm{App}(\mathcal{R})$. The latter of the axioms covers the possibility that $\mathcal{R}$ cannot be applied: in this case, we have an exit condition of $er$ to track its finite failure.

Sequential composition is handled by \textsc{SeqSucc} as well as \textsc{SeqFail} (to cover the possibility of the first program resulting in failure). The conditional construct is covered by \textsc{IfElse}: note that failure can only result from failure in the two branches, and not from the guard $\mathcal{R}$, which is simply tested to choose the branch.

It is important to highlight the rule of consequence, \textsc{Cons}, as the implications in the side conditions are reversed from those of the corresponding Hoare logic rule~\cite{Apt-et_al09a,Hoare69a}. In incorrectness logic, we instead weaken the precondition and strengthen the postcondition. Intuitively, this allows us to soundly drop disjuncts in the result and thus reason about \emph{fewer paths} in the post-state, which may support better scalability in tools~\cite{OHearn20a}.

For the iteration of rule schemata sets, we have a number of cases. The axiom \textsc{IterZero} covers the case when a rule schemata set is no longer applicable (note that this does not result in failure). The proof rule \textsc{Iter} unrolls a step of the iteration. Traditional loop invariants are less important in these proof rules than they are for Hoare logic, as we are reasoning about a subset of paths rather than \emph{all} of them. To see this, consider the triple $\models [inv] \mathcal{R!} [ok: inv \wedge \neg\mathrm{App}(\mathcal{R})]$ with invariant $inv$. Under-approximate validity requires every graph $H$ satisfying $inv$ and $\neg\mathrm{App}(\mathcal{R})$ to be derivable by applying $\mathcal{R!}$ to some graph $G$ satisfying $inv$. One can always find such a graph by taking $G=H$.

Finally, \textsc{IterVar} combines \textsc{IterZero} and \textsc{Iter} into one rule. It expresses that a triple $\vdash [c_0] \mathcal{R!} [ok:c_n]$ can be proven if: (1)~$c_n$ implies the termination of the iteration (i.e.~the non-applicability of $\mathcal{R}$); and (2)~if triples can be proven for the $n$ iterations of $\mathcal{R}$. \textsc{IterVar} is a stricter version of the backwards variant rule for while-loops in~\cite{OHearn20a,Vries-Koutavas11a}: had we adopted the rule in full, we would be able to prove triples such as $\vdash [c(0)] \mathcal{R}! [ok: \exists n. n \geq 0. c(n) \wedge \neg\mathrm{App}(\mathcal{R})]$. Here, $c(i)$ denotes a parameterised predicate, i.e.~in our case, a function mapping expressions to E-constraints. Unfortunately, these are not possible to express using E-constraints, and including them would strictly increase their expressive power beyond first-order graph properties and the current capabilities of `WPost'.

\begin{example}[Colouring: finite failure]\label{ex:finite_failure}
    In our first example, we prove the incorrectness specification $\vdash [ \neg\exists\mathtt{x}.\exists\adjincludegraphics[height=\baselineskip,valign=c]{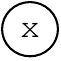} ]\ \mathtt{init}; \mathtt{colour!}\ [ er: \neg\exists\mathtt{x}.\exists\adjincludegraphics[height=\baselineskip,valign=c]{fig_proof_ec_AppInit.pdf} ]$ for the program of Figure~\ref{fig:buggy_program}. This triple specifies that if a graph does not contain any integer-labelled nodes, then it can be derived from another graph satisfying the same condition that the program finitely fails on. Since $\mathtt{init}$ would fail on any such graph, this specification is valid: the graph in the post-state is exactly the graph in the pre-state. Figure~\ref{fig:proof_tree_1} proves this triple using incorrectness logic.

	\begin{figure}[!t]
    	\centering
        \begin{mdframed}
        \begin{prooftree}
            \AxiomC{$\qed$}
            \LeftLabel{\textsc{RuleSetFail}}
            \UnaryInfC{$\vdash [\mathtt{true} \wedge \neg\mathrm{App}(\mathtt{init})]\ \mathtt{init}\ [er : \mathtt{true} \wedge \neg \mathrm{App}(\mathtt{init})]$}
            \LeftLabel{\textsc{Cons}}
    	    \UnaryInfC{$\vdash [\neg \mathrm{App}(\mathtt{init})]\ \mathtt{init}\ [er: \neg \mathrm{App}(\mathtt{init})]$}
     	    \LeftLabel{\textsc{SeqFail}}
        	\UnaryInfC{$\vdash [ \neg \mathrm{App}(\mathtt{init}) ]\ \mathtt{init}; \mathtt{colour!}\ [ er: \neg \mathrm{App}(\mathtt{init}) ]$}
    	\end{prooftree}
    	\end{mdframed}
	    \caption{Proving the presence of failure (E-constraints in Figure~\ref{fig:proof_e-constraints})}\label{fig:proof_tree_1}
	\end{figure}
\end{example}

	\begin{figure}[!t]
    	\centering
        \begin{mdframed}
            \includegraphics[width=\textwidth]{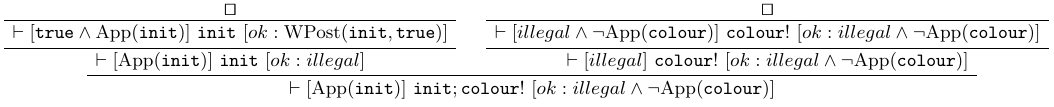}
    	\end{mdframed}
	    \caption{Proving the presence of an illegal graph (E-constraints in Figure~\ref{fig:proof_e-constraints})}\label{fig:proof_tree_iterzero}
	\end{figure}

\begin{figure}[!t]
    \begin{mdframed}
        \centering
        \begin{tabular}{rcl}
            $illegal$ &\ $=$\ & $\left(\exists\mathtt{a},\mathtt{b},\mathtt{j}.\exists \adjincludegraphics[height=11pt,valign=c]{fig_proof_ec_illegal.pdf}\right) \wedge \left(\exists\mathtt{x}.\exists\adjincludegraphics[height=11pt,valign=c]{fig_proof_ec_WPostInit.pdf}\right)$ \\
            $c$ &\ $=$\ & $\exists\mathtt{a}.\exists\adjincludegraphics[height=11pt,valign=c]{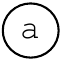}.\neg\exists\mathtt{d},\mathtt{k}.\exists\adjincludegraphics[height=11pt,valign=c]{fig_proof_ec_c.pdf}\adjincludegraphics[height=11pt,valign=c]{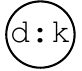}$ \\
            $d$ &\ $=$\ & $\exists\mathtt{a},\mathtt{b},\mathtt{c}.\exists \adjincludegraphics[height=11pt,valign=c]{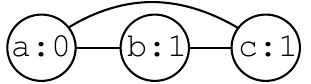}.\neg\exists\mathtt{d},\mathtt{k}.\exists\adjincludegraphics[height=11pt,valign=c]{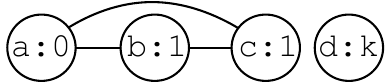}$ \\
            $e$ &\ $=$\ & $\exists\mathtt{a},\mathtt{b},\mathtt{c}.\exists \adjincludegraphics[height=11pt,valign=c]{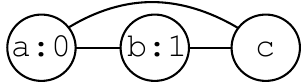}.\neg\exists\mathtt{d},\mathtt{k}.\exists\adjincludegraphics[height=11pt,valign=c]{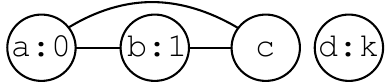}$ \\
            $f$ &\ $=$\ & $\exists\mathtt{a},\mathtt{b},\mathtt{c}.\exists \adjincludegraphics[height=11pt,valign=c]{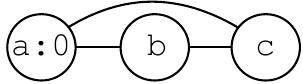}.\neg\exists\mathtt{d},\mathtt{k}.\exists\adjincludegraphics[height=11pt,valign=c]{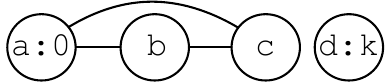}$ \\
            $\mathrm{App}(\mathtt{init})$ &\ $=$\ & $\exists\mathtt{x}.\exists\adjincludegraphics[height=11pt,valign=c]{fig_proof_ec_AppInit.pdf}$ \\
            $\mathrm{App}(\mathtt{colour})$ &\ $=$\ & $\exists\mathtt{x}.\exists\adjincludegraphics[height=11pt,valign=c]{fig_proof_ec_AppColour.pdf}$ \\
            $\mathrm{WPost}(\mathtt{init},\mathtt{true})$ &\ $=$\ & $\exists\mathtt{x}.\exists\adjincludegraphics[height=11pt,valign=c]{fig_proof_ec_WPostInit.pdf}$ \\
            $\mathrm{WPost}(\mathtt{init},c)$ &\ $=$\ & $\exists\mathtt{x}.\exists\adjincludegraphics[height=11pt,valign=c]{fig_proof_ec_WPostInit.pdf}.\left(\neg\exists\mathtt{d},\mathtt{k}.\exists \adjincludegraphics[height=11pt,valign=c]{fig_proof_ec_WPostInit.pdf}\adjincludegraphics[height=11pt,valign=c]{fig_proof_ec_dk.pdf}\right)  \vee \left(\exists\mathtt{a}.\exists\adjincludegraphics[height=11pt,valign=c]{fig_proof_ec_c.pdf}\adjincludegraphics[height=11pt,valign=c]{fig_proof_ec_WPostInit.pdf}.\neg\exists\mathtt{d},\mathtt{k}.\exists\adjincludegraphics[height=11pt,valign=c]{fig_proof_ec_c.pdf}\adjincludegraphics[height=11pt,valign=c]{fig_proof_ec_WPostInit.pdf}\adjincludegraphics[height=11pt,valign=c]{fig_proof_ec_dk.pdf}\right)$ \\
            $\mathrm{WPost}(\mathtt{colour}, e)$ &\ $=$\ & $\left(\exists\mathtt{a,x,y}.\exists \adjincludegraphics[height=11pt,valign=c]{fig_proof_ec_Wpost_e.pdf}.\neg\exists\mathtt{d,k}. \exists\adjincludegraphics[height=11pt,valign=c]{fig_proof_ec_Wpost_e_dk.pdf}\right)$ \\
            && $\vee\ \left(\exists\mathtt{b,x,y}.\exists \adjincludegraphics[height=11pt,valign=c]{fig_proof_ec_Wpost_e2.pdf}.\neg\exists\mathtt{d,k}.\exists\adjincludegraphics[height=11pt,valign=c]{fig_proof_ec_Wpost_e2_dk.pdf} \right) \vee \cdots$ \\
            $\mathrm{WPost}(\mathtt{colour}, f)$ &\ $=$\ & $\left(\exists\mathtt{c,x,y}.\exists \adjincludegraphics[height=11pt,valign=c]{fig_proof_ec_Wpost_f.pdf}.\neg\exists\mathtt{d,k}. \exists\adjincludegraphics[height=11pt,valign=c]{fig_proof_ec_Wpost_f_dk.pdf}\right)$ \\
            && $\vee\ \left(\exists\mathtt{b,c,x,y}.\exists \adjincludegraphics[height=11pt,valign=c]{fig_proof_ec_Wpost_f2.pdf}.\neg\exists\mathtt{d,k}.\exists\adjincludegraphics[height=11pt,valign=c]{fig_proof_ec_Wpost_f2_dk.pdf} \right) \vee \cdots$
        \end{tabular}
    \end{mdframed}
    \caption{E-constraints used in the proofs of Figures~\ref{fig:proof_tree_1}, \ref{fig:proof_tree_iterzero}, and \ref{fig:proof_tree_2}}
    \label{fig:proof_e-constraints}
\end{figure}

\begin{sidewaysfigure}
    		\centering
            \begin{mdframed}
                \includegraphics[width=\textwidth]{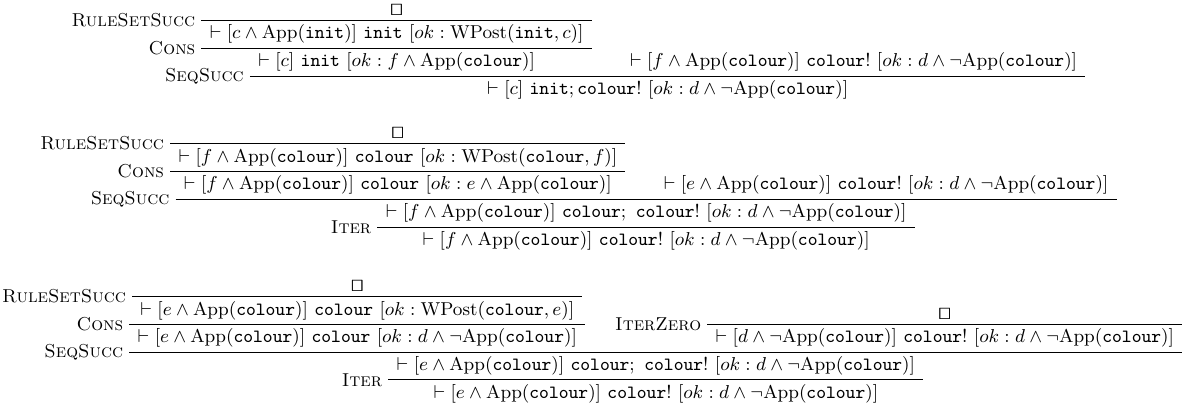}
        	\end{mdframed}
    	    \caption{Proving the presence of an illegal colouring (E-constraints in Figure~\ref{fig:proof_e-constraints})}\label{fig:proof_tree_2}
	    \end{sidewaysfigure}

\begin{example}[Colouring: illegal graph]\label{ex:illegal_graph}
    While proving the presence of failure for the program of Figure~\ref{fig:buggy_program} is simple, there are some interesting subtleties involved in proving the presence of illegal graph structure. Let us consider:
    \[ \vdash [ \exists\mathtt{x}.\exists\adjincludegraphics[height=\baselineskip,valign=c]{fig_proof_ec_AppInit.pdf} ]\ \mathtt{init}; \mathtt{colour!}\ [ ok: \left(\exists\mathtt{a},\mathtt{b},\mathtt{j}.\exists \adjincludegraphics[height=\baselineskip,valign=c]{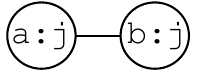}\right) \wedge \left(\exists\mathtt{x}.\exists\adjincludegraphics[height=11pt,valign=c]{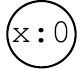}\right) \wedge \left(\neg\exists\mathtt{x}.\exists\adjincludegraphics[height=\baselineskip,valign=c]{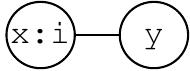}\right) ] \]
    
    \noindent which specifies that if a graph has an illegal colouring, at least one node coloured `0', and $\mathtt{colouring}$ is no longer applicable, then it can be derived by applying the program to some graph containing an integer-labelled node (i.e.~that $\mathtt{init}$ does not fail on). This triple is provable (Figure~\ref{fig:proof_tree_iterzero}) and valid, but not because of any problem with $\mathtt{colour}$. Consider, for example, the graph \adjincludegraphics[height=\baselineskip,valign=c]{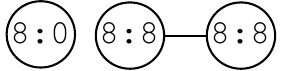}. This is trivially reachable from graphs that already contain the illegal structure, e.g.~\adjincludegraphics[height=\baselineskip,valign=c]{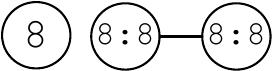}, thus we are able to complete the proof using the \textsc{IterZero} rule.

    Finally, we strengthen the condition on the result to try and prove the presence of an illegal colouring that is created by the program itself (see Figure~\ref{fig:proof_e-constraints} for the E-constraints):
        \[ \vdash [ c ]\ \mathtt{init}; \mathtt{colour!}\ [ ok: d \wedge \neg\mathrm{App}(\mathtt{colour})  ] \]
    
    \noindent The E-constraint $c$ expresses that there exists at least one node and that no node is coloured (instead of using conjunction, we express this more compactly using nesting). The E-constraint $d$ expresses that there are three coloured nodes (with colours $\mathtt{0,1,1}$). Together, the triple specifies that every graph satisfying $d \wedge \neg\mathrm{App}(\mathtt{colour})$ can be derived from at least one graph satisfying $c$. This triple is valid and provable (Figure~\ref{fig:proof_tree_2}) as the illegal colouring is a logical possibility of some executions of $\mathtt{colour!}$. Note that we cannot use an assertion such as $\exists\mathtt{a},\mathtt{b}.\exists \adjincludegraphics[height=\baselineskip,valign=c]{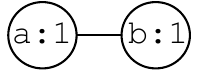}$ in place of $d$, as this is satisfied by the graph \adjincludegraphics[height=\baselineskip,valign=c]{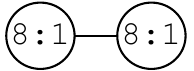} which is impossible to derive from any graph satisfying $c$.
    
    As E-constraints are equivalent to first-order logic on graphs~\cite{Poskitt13a}, we are precluded from proving a more general non-local condition, e.g.~``there exists a cycle with an illegal colouring''. However, there are more powerful logics equipped with similar transformations that may be possible to use instead~\cite{Navarro-et_al21a,Poskitt-Plump14a}.

\end{example}

\section{Transformations, Soundness, and Completeness}
\label{sec:results}

This section presents formal definitions and characterisations of the transformations that are used in some of our incorrectness axioms and proof rules. Following this, we present our main technical result: the soundness and completeness of our incorrectness logic with respect to the denotational semantics.

First, we consider `App', which transforms a set of rule schemata into an E-constraint that expresses the minimum requirements on a graph for at least one of the rules to be applicable. Intuitively, the E-constraint expresses the presence of a match for a left-hand side, i.e.~a morphism that satisfies the dangling condition. This transformation is adapted from similar transformations in~\cite{Habel-Pennemann09a,Poskitt13a}.

\begin{proposition}[Applicability]\label{prop:app}\rm
    For every graph $G\in\mathcal{G}(\mathcal{L})$ and set of rule schemata $\mathcal{R}$,
    \[ G \models \mathrm{App}(\mathcal{R})\ \mathrm{if\ and\ only\ if}\ \exists H.\ G\Rightarrow_\mathcal{R} H.\]
    
    \noindent\emph{Construction}. Define $\mathrm{App}(\emptyset) = \mathtt{false}$ and then $\mathrm{App}(\{r_1, \cdots r_n\}) = \mathrm{app}(r_1) \vee \cdots \mathrm{app}(r_n)$. Given a rule schema $r = \langle L \hookleftarrow K \hookrightarrow R \rangle$ over variables $\mathtt{x}_1,\cdots\mathtt{x}_m$, define $\mathrm{app}(r) = \exists\mathtt{x}_1,\cdots\mathtt{x}_m.\ \exists\emptyset\hookrightarrow L.\ \mathrm{Dang}(r)$.
    
    Finally, define $\mathrm{Dang}(r) = \bigwedge_{a\in A} \neg \exists \mathtt{x}_a.\exists a$ where the index set $A$ ranges over all injective morphisms (equated up to isomorphic codomains) $a\!: L \hookrightarrow L^\oplus$ such that the pair $\langle K\hookrightarrow L, a \rangle$ has no natural pushout complement and each $L^\oplus$ is a graph that can be obtained from $L$ by adding either: (1)~a single loop with label $\square$; (2)~a single edge with label $\square$ between distinct nodes; or (3)~a single node labelled with fresh variable $\mathtt{x}_a$ and a non-looping edge incident to it with label $\square$. If the index set $A$ is empty, then $\mathrm{Dang}(r) = \mathtt{true}$.
    \qed
\end{proposition}

Next, we consider `WPost', which transforms a set of rule schemata and a presumption into a weakest postcondition, i.e.~the weakest property a graph must satisfy to guarantee the \emph{existence} of a pre-state that satisfies the presumption. WPost is defined via two intermediate transformations: `Shift' and `Right'.

We begin by defining `Shift', which can be used to transform an E-constraint $c$ into an E-condition over the left-hand side of a rule $L$ by considering all the ways that a `match' can overlap with $c$. Our definition is adapted from the shifting constructions of~\cite{Habel-Pennemann09a,Poskitt13a} to handle the explicit quantification of label variables. Intuitively, this step is handled via a disjunction over all possible substitutions of a variable in $c$ for integer expressions or variables in $L$, i.e.~to account for interpretations in which they refer to the same values.

To facilitate this, we require that the labels in $c$ are lists of variables that are distinct from those in $L$. This is a mild assumption, as an arbitrary expression can simply be replaced with a variable that is then equated with the original expression in an interpretation constraint.

\begin{lemma}[E-constraint to left E-condition]\label{lemma:shift}\rm
    Let $r$ denote a rule schema and $c$ an E-constraint labelled over lists of variables distinct from those in $r$. For every graph $G\in\mathcal{G}(\mathcal{L})$ and morphism $g\!: L^I \hookrightarrow G$ with $\mathrm{dom}(I) = \mathrm{vars}(L)$,
    \[ g\!: L^I \hookrightarrow G \models^I \mathrm{Shift}(r,c) \ \mathrm{if\ and\ only\ if}\ G \models c. \]
\end{lemma}
    
    \noindent \emph{Construction.} Let $c$ denote an E-constraint and $r$ a rule with left-hand side $L$. We define $\mathrm{Shift}(r,c) = \mathrm{Shift}'(\emptyset\hookrightarrow L, c)$. We define $\mathrm{Shift}'$ inductively for morphisms $p\!:P\hookrightarrow P'$ and E-conditions over $P$. Let $\mathrm{Shift}'(p, \mathtt{true}) = \mathtt{true}$ and $\mathrm{Shift}'(p,\gamma) = \gamma$. Then:
    
    \begin{equation*}
        \begin{split}
            \mathrm{Shift}'(p, \exists\mathtt{x}.\ c) & = \left(\exists\mathtt{x}.\ \mathrm{Shift}'(p,c)\right) \bigvee_{l\in\Sigma_{P'}} \mathrm{Shift}'(p,c^{(\mathtt{x}\mapsto l)}) \\
            \mathrm{Shift}'(p, \exists a\!:P\hookrightarrow C.\ c) & = \bigvee_{e\in\varepsilon} \exists b\!: P' \hookrightarrow E.\ \mathrm{Shift}'(s\!:C\hookrightarrow E, c)
        \end{split}
    \end{equation*}
    
    \begin{wrapfigure}[10]{r}{0.25\textwidth}
	\centering
		\vspace{-25pt}
		\includegraphics[width=0.225\textwidth]{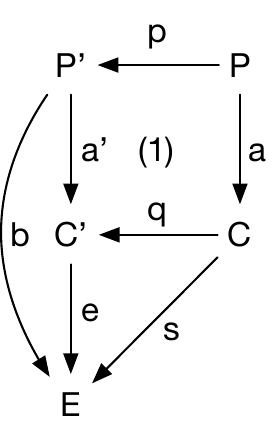}
    \end{wrapfigure}
    
    \noindent In the third case, $\Sigma_{P'}$ is the set of all variables and integer expressions present in the labels of $V_{P'}$. In the fourth case, construct pushout (1) of $p$ and $a$ as depicted in the diagram. The disjunction ranges over the set $\varepsilon$, which we define to contain every surjective morphism $e\!: C'\hookrightarrow E$ such that $b = e \circ a'$ and $s = e \circ q$ are injective morphisms. (We consider codomains of each $e$ up to isomorphism, so the disjunction is finite.)
    
    Shift and Shift' are defined for Boolean formulae over E-conditions in the standard way.
    \qed

\begin{example}[Shift]\label{ex:shift}
    Consider the rule schema $\mathtt{init}$ (Figure~\ref{fig:buggy_program}) and E-constraint $c$ (Figure~\ref{fig:proof_e-constraints}). After simplification, the transformation $\mathrm{Shift}(\mathtt{init},c)$ results in:\\
    
    \noindent\includegraphics[width=\textwidth]{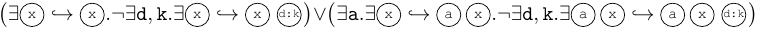}

\end{example}

The second intermediate transformation for `WPost' is `Right', which transforms an E-condition over the left-hand side of a rule to an E-condition over the right-hand side. This construction is based on transformation `L' from~\cite{Habel-Pennemann09a,Poskitt13a} but in the reverse direction.

\begin{lemma}[Left to right E-condition]\label{lemma:right}\rm
    Let $r = \langle L \hookleftarrow K \hookrightarrow R\rangle$ denote a rule schema and $c$ an E-condition over $L$. Then for every direct derivation $G\Rightarrow_{r,g,h} H$ with $g\!:L^I\hookrightarrow G$ and $h\!:R^I\hookrightarrow H$,
    \[ g\!: L^I \hookrightarrow G \models^I c\ \mathrm{if\ and\ only\ if}\ h\!:R^I\hookrightarrow H \models^I \mathrm{Right}(r,c). \]
\end{lemma}

    \begin{wrapfigure}{r}{0.24\textwidth}
	\centering
		\vspace{10pt}
		\includegraphics[width=0.23\textwidth]{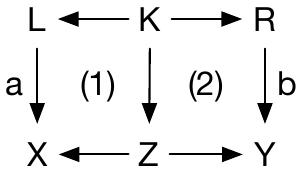}
    \end{wrapfigure}

    \noindent \emph{Construction.} We define $\mathrm{Right}(r,\mathtt{true}) = \mathtt{true}$, $\mathrm{Right}(r,\gamma) = \gamma$, and $\mathrm{Right}(r,\exists\mathtt{x}.\ c) = \exists\mathtt{x}.\ \mathrm{Right}(r,c)$. Let $\mathrm{Right}(r,\exists a.\ c) = \exists b.\ \mathrm{Right}(r^\ast,c)$ if $\langle K \hookrightarrow L, a \rangle$ has a natural pushout complement (1), where $r^\ast = \langle X \hookleftarrow Z \hookrightarrow Y \rangle$ denotes the rule `derived' by also constructing natural pushout (2). If $\langle K \hookrightarrow L, a \rangle$ has no natural pushout complement, then $\mathrm{Right}(r, \exists a.\ c) = \mathtt{false}$.
    
    Right is defined for Boolean formulae over E-conditions as per usual.
    \qed

\begin{example}[Right]\label{ex:right}
    Continuing from Example~\ref{ex:shift}, applying the transformation $\mathrm{Right}(\mathtt{init}, \mathrm{Shift}(\mathtt{init}, c) )$ results in the E-condition:\\
    
    \noindent\includegraphics[width=\textwidth]{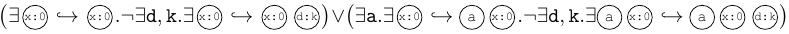}

\end{example}

Next, we can give `WPost' a simple definition based on the two intermediate transformations. Intuitively, it constructs a disjunction of E-constraints that demand the existence of some co-match that would result from applying the rule schema set to a graph satisfying the presumption.

\begin{proposition}[Weakest postcondition]\label{prop:wpost}\rm
    Let $\mathcal{R}$ denote a rule schemata set and $c$ an E-constraint. Then for every graph $H\in\mathcal{G}(\mathcal{L})$,
    \[ H \models \mathrm{WPost}(\mathcal{R}, c)\  \text{if and only if}\  \exists G.\ G \models c\ \mathrm{and}\ G\Rightarrow_\mathcal{R} H. \]
    
    \noindent \emph{Construction.} Define $\mathrm{WPost}(\emptyset,c) = \mathtt{false}$ and  $\mathrm{WPost}(\mathcal{R},c) = \bigvee_{r\in\mathcal{R}}\mathrm{wpost}(r,c)$. Let $\mathrm{wpost}(r,c) = \exists\mathtt{x}_1,\cdots\mathtt{x}_n.\exists \emptyset \hookrightarrow R.\mathrm{Dang}(r^{-1})\wedge\mathrm{Right}(r, \mathrm{Shift}(r, c) )$ where $\{\mathtt{x}_1, \cdots ,\mathtt{x}_n \} = \mathrm{vars}(R)$ and $r^{-1}$ is the reversal of rule $r$.
    \qed
\end{proposition}

\begin{example}[WPost]
    Continuing from Example~\ref{ex:right}, applying the transformation $\mathrm{WPost}(\mathtt{init},c)$ results in the E-constraint given in Figure~\ref{fig:proof_e-constraints}.
\end{example}

Finally, using the characterisations of `App' and `WPost', we can present the main technical results of our paper: the soundness and completeness of our incorrectness logic for graph programs. Soundness means that any triple provable in our logic is valid in the sense of Definition~\ref{defn:validity}, i.e.~that graphs satisfying the result are reachable from some graph satisfying the presumption. The proof of this theorem is by structural induction on triples.

\begin{theorem}[Soundness]\label{thm:soundness}\rm
    For all E-constraints $c,d$, graph programs $P$, and exit conditions $\epsilon$,
    \[ \vdash [c]\ P\ [\epsilon: d]\ \mathrm{implies}\ \models [c]\ P\ [\epsilon: d]. \]
    \qed
\end{theorem}

Completeness is the other side of the coin: it means that any valid triple can be proven using our logic. As is typical, we prove \emph{relative completeness}~\cite{Cook78a} in which completeness is relative to the existence of an oracle for deciding the validity of assertions (as in \textsc{Cons}). The idea is to separate incompleteness due to the incorrectness logic from incompleteness in deducing valid assertions, and determine that no proof rules are missing. Our proof relies on some semantically (or extensionally) defined assertions, $\mathrm{WPOST}[P,c]$, that characterise exactly the weakest postcondition of an arbitrary program $P$ relative to an E-constraint $c$.

\begin{theorem}[Relative completeness]\label{thm:completeness}\rm
    For all E-constraints $c,d$, graph programs $P$, and exit conditions $\epsilon$,
    \[ \models [c]\ P\ [\epsilon: d]\ \mathrm{implies}\ \vdash [c]\ P\ [\epsilon: d]. \]
    \qed
\end{theorem}

It is important to remark that it is unknown whether E-constraints are \emph{expressive} enough to specify precisely the assertion $\mathrm{WPOST}[P,c]$ in general; in fact, there is evidence to suggest they may not be~\cite{Wulandari-Plump20a}. This is, however, a limitation of the logic and not the incorrectness proof rules, and expressiveness may not be a problem faced by stronger assertion languages for graphs, such as those supporting non-local properties~\cite{Navarro-et_al21a,Orejas-et_al18a,Poskitt-Plump14a}.

\section{Related Work}
\label{sec:related_work}

Over-approximate program logics for proving the absence of bugs have been studied extensively~\cite{Apt-et_al09a}. Our program logic differs by focusing on under-approximate reasoning, i.e.~proofs about the presence of bugs (in our case, forbidden graph structure or finitely failing execution paths). The first under-approximate calculus of this kind was introduced by De Vries and Koutavas~\cite{Vries-Koutavas11a}, who proposed the notion of under-approximate validity, and defined a `Reverse Hoare Logic' for proving reachability specifications over the proper states of imperative randomised programs. O'Hearn's incorrectness logic~\cite{OHearn20a} extended this program logic to support under-approximate reasoning about executions that result in errors, an idea we adopt to support reasoning about both successful computations ($ok$) and finitely failing executions ($er$). Both of these program logics use variants to reason about while-loop termination, but unlike standard Hoare logics, require that the variant decreases in the backwards direction. Our \textsc{IterVar} rule is similar, but requires the number of iterations to be known as E-conditions are not expressive enough to specify parameterised graph properties, for example, the existence of a cycle of length $n$.

Raad et al.~\cite{Raad-et_al20a} combined separation logic with incorrectness logic to facilitate proofs about the presence of bugs using local reasoning, i.e.~specifications that focus only on the region of memory being accessed. They found that the original model of separation logic, which does not distinguish dangling pointers from pointers we have no knowledge about, to be incompatible with the under-approximate frame rule. This was resolved by refining the model with negative heap assertions that can specify that a location has been de-allocated. 

Murray~\cite{Murray20a} proposed the first under-approximate relational logic, allowing for reasoning about the behaviours of pairs of programs. As many important security properties (e.g.~noninterference, function sensitivity, refinement) can be specified as relational properties, Murray's program logic can be used to provably demonstrate the presence of insecurity.

Bruni et al.~\cite{Bruni-et_al21a} incorporate incorrectness logic in a proof system for abstract interpretation that combines over- and under-approximation. Given an abstraction that is `locally complete' (i.e.~complete only for some specific inputs, rather than all possible inputs), they show that it is possible to prove both the presence as well as the absence of true alerts.

Incorrectness logics allow formal reasoning about reachability specifications---in our context, the presence of finite failure or forbidden graph structure. A complementary approach is to find counterexamples (i.e.~instances of the forbidden structure) using model checkers such as \textsc{Groove}~\cite{Ghamarian-et_al12a}. Analysing graph transformation systems can be challenging, however, as they often have infinite state spaces, but this can be mitigated by using bounded model checking~\cite{Isenberg-Steenken-Wehrheim13a}.

\section{Conclusion and Future Work}
\label{sec:conclusion}

We proposed an incorrectness logic for under-approximate reasoning about graph programs, demonstrating that the deductive rules of Hoare logics can be `reversed' to prove the presence of graph transformation bugs, such as the possibility of illegal graph substructures or finitely failing execution paths. In particular, we presented a calculus of incorrectness axioms and rules, proved them to be sound and relatively complete with respect to a denotational semantics of graph programs, and demonstrated their use to prove the presence of various bugs in a faulty node colouring program.

This paper was principally a theoretical exposition, but was motivated by some potentially interesting applications. One idea (suggested by O'Hearn~\cite{OHearn20a}) is to recast static bug catchers in terms of finding under-approximation proofs. For instance, incorrectness logic might be able to provide soundness arguments for various approaches that symbolically execute graph or model transformations (e.g.~\cite{AlSibahi-Dimovski-Wasowski16a,Azizi-et_al20a,Oakes-et_al18a}). Another idea is to use it to complement over-approximate proofs: if one is unable to prove a partial correctness specification or the absence of failure~\cite{Poskitt-Plump13a}, switch to under-approximate proofs instead and reason about the circumstances that could cause some undesirable result to be reachable.

Beyond exploring these potential applications, future work should also extend our logic to a full-fledged graph programming language (e.g.~\textsc{GP 2}~\cite{Plump11a}, or the recipes of \textsc{Groove}~\cite{Corrodi-Heussner-Poskitt18a,Ghamarian-et_al12a}). It is also important to investigate how to make incorrectness reasoning for graph programs easier. This could be in the form of guidelines on how to come up with incorrectness specifications (reasoning over a whole graph can be counter-intuitive, as Examples~\ref{ex:finite_failure} and \ref{ex:illegal_graph} demonstrate), or some derived proof rules for simplifying reasoning about common patterns.\\

\noindent \textbf{Acknowledgements.} I am grateful to the ICGT'21 referees for their detailed reviews and suggestions, which have helped to improve the quality of this paper.

\bibliographystyle{splncs04}
\bibliography{references}


\section*{Appendix}

\begin{proof}[Proposition~\ref{prop:app}; Lemmata~\ref{lemma:shift}--\ref{lemma:right}]
    By induction over the form of E-conditions, following the proof structure for transformations `App', `A', and `L' for the similar assertion language in~\cite{Poskitt13a}.
    \qed
\end{proof}

\begin{proof}[Proposition~\ref{prop:wpost}]
    $\Longrightarrow$. Assume that $H\models \mathrm{WPost}(\mathcal{R},c)$. There exists some $r\in\mathcal{R}$ such that:
    \[ H \models \mathrm{wpost}(r,c) = \exists\mathtt{x}_1,\cdots\mathtt{x}_n.\exists \emptyset \hookrightarrow R.\mathrm{Dang}(r^{-1})\wedge\mathrm{Right}(r, \mathrm{Shift}(r, c) ). \]
    
    \noindent There exists an $h\!: R^I\hookrightarrow G$ such that $h \models^I \mathrm{Dang}(r^{-1})\wedge\mathrm{Right}(r, \mathrm{Shift}(r, c) )$. Using Proposition~\ref{prop:app}, there exists a direct derivation from some graph $G$ to $H$ via $r = \langle L\Rightarrow R \rangle$, and by Lemma~\ref{lemma:right}, there exists some $g\!:L^I\hookrightarrow G$ such that $g \models^I \mathrm{Shift}(r,c)$. By Lemma~\ref{lemma:shift}, $G\models c$.
    
    $\Longleftarrow$. Assume that there exists a graph $G$ such that $G\models c$ and $G\Rightarrow_\mathcal{R} H$. There exists some $r = \langle L \Rightarrow R \rangle \in \mathcal{R}$ such that $G\Rightarrow_r H$. By the definition of $\models$, Lemma~\ref{lemma:shift}, and Lemma~\ref{lemma:right}, there exists some $h\!:R^I\hookrightarrow G \models^I \mathrm{Right}(r,\mathrm{Shift}(r,c))$. By the definition of direct derivations and Proposition~\ref{prop:app}, $h \models^I \mathrm{Dang}(r^{-1})$, and thus $h \models^I \mathrm{Dang}(r^{-1}) \wedge \mathrm{Right}(r,\mathrm{Shift}(r,c))$. By the definition of $\models$, $H \models \exists\mathtt{x}_1,\cdots\mathtt{x}_n.\exists R.\mathrm{Dang}(r^{-1}) \wedge \mathrm{Right}(r,\mathrm{Shift}(r,c))$, that is, $H \models \mathrm{wpost}(r,c)$. Being a disjunct of $\mathrm{WPost}(r,c)$, we derive the result $H \models \mathrm{WPost}(r,c)$.
    \qed
\end{proof}

\begin{proof}[Theorem~\ref{thm:soundness}]
    Given $\vdash [c] P [\epsilon: d]$, we need to show that $\models [c] P [\epsilon: d]$. We consider each axiom and proof rule in turn and proceed by induction on proofs.
    
    \textsc{RuleSetSucc}, \textsc{RuleSetFail}. The validity of these axioms follows immediately from the definitions of $\llbracket \mathcal{R} \rrbracket ok$, $\llbracket \mathcal{R} \rrbracket er$, Proposition~\ref{prop:app}, and Proposition~\ref{prop:wpost}.
    
    \textsc{SeqSucc}. Suppose that $\vdash [c] P;Q [ok: d]$. By induction, we have $\models [c] P [ok: e]$ and $\models [e] Q [ok: d]$. By definition of $\models$, for all $H. H \models d$, there exists a $G'.G'\models e$ with $(G',G) \in \llbracket Q \rrbracket ok$, and for all $G'. G' \models e$, there exists a $G.G \models c$ with $(G,G') \in \llbracket P \rrbracket ok$. From the definition of $\models$ and $\llbracket P;Q \rrbracket ok$, it then follows that $\models [c] P;Q [ok: d]$. Analogous for case $\vdash [c] P;Q [er: d]$.
    
    \textsc{SeqFail}. Suppose that $\vdash [c] P;Q [er: d]$. By induction, we have $\models [c] P [er: d]$. By definition of $\models$, for all $H. H \models d$, there exists a $G. G \models c$ with $(G,H) \in \llbracket P \rrbracket er$. By the definition of $\llbracket P;Q \rrbracket er$ and $\models$, it follows that $\models [c] P;Q [er: d]$.
    
    \textsc{IfElse}. Suppose that $\vdash [c] \mathtt{if}\ \mathcal{R}\ \mathtt{then}\ P\ \mathtt{else}\ Q [\epsilon: d]$. By induction, we have $\models [c \wedge\mathrm{App}(\mathcal{R})] P [\epsilon: d]$ or $\models [c \wedge \neg\mathrm{App}(\mathcal{R})] Q [\epsilon: d]$. From the definition of $\models$, $\llbracket \mathtt{if}\ \mathcal{R}\ \mathtt{then}\ P\ \mathtt{else}\ Q \rrbracket \epsilon$, and Proposition~\ref{prop:app}, we obtain the result that $\models [c] \mathtt{if}\ \mathcal{R}\ \mathtt{then}\ P\ \mathtt{else}\ Q [\epsilon: d]$.
    
    \textsc{Cons}. Suppose that $\vdash [c] P [\epsilon: d]$. By induction, we have $\models [c'] P [\epsilon: d']$, $\models d \Longrightarrow d'$, and $\models c' \Longrightarrow c$. It immediately follows that $\models [c] P [\epsilon: d]$.
    
    \textsc{IterZero}. For every graph $G. G\models c \wedge \neg\mathrm{App}(\mathcal{R})$, by Proposition~\ref{prop:app}, $G\not\Rightarrow_\mathcal{R}$, $(G,G)\in\llbracket \mathcal{R}\rrbracket er$, and thus $(G,G) \in \llbracket \mathcal{R}! \rrbracket ok$. It immediately follows that $\models [c \wedge \neg\mathrm{App}(\mathcal{R})] \mathcal{R}! [ok: c \wedge \neg\mathrm{App}(\mathcal{R})]$.
    
    \textsc{Iter}. Suppose that $\vdash [c \wedge \mathrm{App}(\mathcal{R})] \mathcal{R!} [ok: d \wedge \neg\mathrm{App}(\mathcal{R})]$. By induction, $\models [c \wedge \mathrm{App}(\mathcal{R})] \mathcal{R;R!} [ok: d \wedge \neg\mathrm{App}(\mathcal{R})$. By definition of $\models$, for all $H. H \models d \wedge \neg\mathrm{App}(\mathcal{R})$, there exists some $G. G\models c \wedge \mathrm{App}(\mathcal{R})$ and $(G,H) \in \llbracket \mathcal{R};\mathcal{R}! \rrbracket ok$. By the definition of $\llbracket \mathcal{R}! \rrbracket ok$ and $\models$, we obtain $\models [c \wedge \mathrm{App}(\mathcal{R})] \mathcal{R!} [ok: d \wedge \neg\mathrm{App}(\mathcal{R})]$.

    \textsc{IterVar}. Suppose that $\vdash [c_0] \mathcal{R}! [ok: c_n]$. By induction, $\models [c_{i-1}] \mathcal{R} [ok: c_i]$ for every $0 < i \leq n$ and $\models c_n \Longrightarrow \neg\mathrm{App}(\mathcal{R})$. By the definition of $\models$ and $\llbracket \mathcal{R} \rrbracket ok$, for every $G_i. G_i \models c_i$, there exists some $G_{i-1}. G_{i-1} \models c_{i-1}$ and $G_{i-1} \Rightarrow_\mathcal{R} G_i$. It follow that there is a sequence of derivations $G_0 \Rightarrow_\mathcal{R} \cdots \Rightarrow_\mathcal{R} G_n$ with $G_0 \models c_0$ and $G_n \models c_n$. By $\models c_n \Longrightarrow \neg\mathrm{App}(\mathcal{R})$ and Proposition~\ref{prop:app}, we have $G_n \not\Rightarrow_\mathcal{R}$, i.e.~$(G_n,G_n) \in \llbracket \mathcal{R} \rrbracket er$. Together with the definition of $\llbracket \mathcal{R}! \rrbracket ok$, it follows that $\models [c_0] \mathcal{R}! [ok: c_n]$.
    \qed
\end{proof}

\begin{proof}[Theorem~\ref{thm:completeness}]
    We prove relative completeness extensionally by showing that for every program $P$, extensional assertion $c$, and exit condition $\epsilon \in \{ok, er\}$, $\vdash [c]P[\epsilon: \mathrm{WPOST}[P,c]]$, where $\mathrm{WPOST}[P,c]$ is an extensional assertion expressing the weakest postcondition relative to $P$ and $c$, i.e.~if $\models [c] P [\epsilon: d]$ for any $d$, then $d\Longrightarrow \mathrm{WPOST}[P,c]$ is valid. Relative completeness is obtained by applying the rule of consequence to $\vdash [c]P[\epsilon: \mathrm{WPOST}[P,c]]$.

    \emph{Rule Application ($\epsilon = ok$).} Immediate from \textsc{RuleSetSucc} and \textsc{Cons}.
    
    \emph{Rule Application ($\epsilon = er$).} Immediate from \textsc{RuleSetFail}, the definition of $\llbracket \mathcal{R} \rrbracket er$, and \textsc{Cons}.

    \emph{Sequential Composition ($\epsilon=ok$).} In this case,
    \begin{align*}
         & H \models \mathrm{WPOST}[P;Q,c] \\
         \mathrm{iff}\ & \exists G. G\models c\ \text{and}\ (G,H) \in \llbracket P;Q \rrbracket ok\\
         \mathrm{iff}\ & \exists G,G'. G\models c, (G,G')\in\llbracket P \rrbracket ok,\ \text{and}\ (G',H)\in \llbracket Q \rrbracket ok\\
         \mathrm{iff}\ & \exists G'. G'\models \mathrm{WPOST}[P,c]\ \text{and}\ (G',H)\in \llbracket Q \rrbracket ok\\
         \mathrm{iff}\ & H \models \mathrm{WPOST}[Q,\mathrm{WPOST}[P,c]]
    \end{align*}
    
    \noindent By induction we have $\vdash [\mathrm{WPOST}[P,c]] Q [ok: \mathrm{WPOST}[Q,\mathrm{WPOST}[P,c]]]$ and $\vdash [c] P [ok: \mathrm{WPOST}[P,c]]$. By \textsc{SeqSucc} we derive the triple $\vdash [c] P;Q [ok: \mathrm{WPOST}[Q,\mathrm{WPOST}[P,c]]]$, and by \textsc{Cons} $\vdash [c] P;Q [ok: \mathrm{WPOST}[P;Q,c]]$.
    
    \emph{Sequential Composition ($\epsilon=er$).} If the program $P;Q$ fails and the error occurs in $Q$, then the proof is analogous to the $ok$ case. If the error occurs in $P$:
        \begin{align*}
         & H \models \mathrm{WPOST}[P;Q,c]\\
         \mathrm{iff}\ & \exists G. G\models c\ \text{and}\ (G,H) \in \llbracket P;Q \rrbracket er\\
         \mathrm{iff}\ & \exists G. G\models c\ \text{and}\ (G,H) \in \llbracket P \rrbracket er\\
         \mathrm{iff}\ & H \models \mathrm{WPOST}[P,c]
    \end{align*}
    
    \noindent By induction we have $\vdash [c] P [ er: \mathrm{WPOST}[P,c]]$, and by \textsc{SeqFail} derive $\vdash [c] P;Q [er: \mathrm{WPOST}[P,c]]$. With \textsc{Cons} we get $\vdash [c] P;Q [er: \mathrm{WPOST}[P;Q,c]]$.
    
    \emph{If-then-else.} The proof for this case follows a similar structure to sequential composition but treating the two branches separately.
    
    \emph{Iteration.} Define $c_i$ as $\mathrm{WPOST}[\mathcal{R},c_{i-1}]$ for every $0 < i \leq n$. We have:
    \begin{align*}
        & G_n \models \mathrm{WPOST}[\mathcal{R!},c_0]\\
        \mathrm{iff}\ & \exists G_0. G_0 \models c_0\ \text{and}\ (G_0,G_n) \in \llbracket \mathcal{R!} \rrbracket ok\\
        \mathrm{iff}\ & \exists G_0,\cdots G_{n-1}. (G_{i-1},G_i) \in \llbracket \mathcal{R} \rrbracket ok\ \text{for all}\ 0 < i \leq n,\ \text{and}\ (G_n, G_n) \in \llbracket \mathcal{R} \rrbracket er \\
        \mathrm{iff}\ & \exists G_1,\cdots G_{n-1}. G_1 \models \mathrm{WPOST}[\mathcal{R},c_0],\ (G_{i-1},G_i) \in \llbracket \mathcal{R} \rrbracket ok\ \text{for all}\ 1 < i \leq n\\
        &\ \ \text{and}\ (G_n, G_n) \in \llbracket \mathcal{R} \rrbracket er \\
        \mathrm{iff}\ & G_n \models c_n\ \text{and}\ c_n \Longrightarrow \neg\mathrm{App}(\mathcal{R})
    \end{align*}
    \noindent By induction,  $\vdash [c_{i-1}] \mathcal{R} [ok: \mathrm{WPOST}[\mathcal{R},c_{i-1}]]$ and thus $\vdash [c_{i-1}] \mathcal{R} [ok: c_i]$. By \textsc{IterVar} and \textsc{Cons} derive the result, $\vdash [c_0] \mathcal{R!} [ok: \mathrm{WPOST}[\mathcal{R!},c_0]]$.
    \qed
\end{proof}

\end{document}